 \documentclass[twocolumn]{aastex631}
\usepackage{amsmath}
\usepackage{breqn}
\usepackage{comment}

\font\manual=manfnt at 7pt \def\dbend{\hbox{\raise0.9ex\hbox{\manual\char127\hspace{0.6em}}}}

\providecommand{\e}[1]{\ensuremath{\times 10^{#1}}}

\newcounter{INTERNALionstage}

\def\gtsim{\mathrel{\hbox{\rlap{\hbox{\lower4pt\hbox{$\sim$}}}\hbox{$>$}}}}
\def\lesssim{\mathrel{\hbox{\rlap{\hbox{\lower4pt\hbox{$\sim$}}}\hbox{$<$}}}}

\def\A{{\rm\thinspace \AA}}

\def\g{{\rm\thinspace g}}

\def\K{{\rm\thinspace K}}

\def\m{{\rm\thinspace m}}

\def\s{{\rm\thinspace s}}

\def\W{{\rm\thinspace W}}

\def\h0{\mbox{{\rm H}$^0$}}
\DeclareMathAlphabet{\vib}{OML}{cmm}{m}{it}
\newcommand*{\satellite}[1]{\textit{#1}}

\newcommand*{\chandra}{\satellite{Chandra}}

\renewcommand{\thefootnote}{\fnsymbol{footnote}}

\shorttitle{Particle acceleration in Abell 1240}
\graphicspath{{./}{figures/}}
\begin{document}

\title{
On the Particle Acceleration Mechanisms in a Double Radio Relic Galaxy Cluster, Abell 1240}

\author[0000-0002-5222-1337]{Arnab Sarkar}
\affiliation{Kavli Institute for Astrophysics and Space Research,
Massachusetts Institute of Technology, 70 Vassar St, Cambridge, MA 02139}
\email{arnabsar@mit.edu}

\author[0000-0002-8144-9285]{Felipe Andrade-Santos}
\affiliation{Department of Liberal Arts and Sciences, Berklee College of Music, 7 Haviland Street, Boston, MA 02215, USA}
\affiliation{Center for Astrophysics $\vert$ Harvard \& Smithsonian, Cambridge, MA 02138, USA}

\author[0000-0002-0587-1660]{Reinout J. van Weeren}
\affiliation{Leiden Observatory, Leiden University, PO Box 9513, 2300 RA Leiden, The Netherlands}

\author[0000-0002-0765-0511]{Ralph P. Kraft}
\affiliation{Center for Astrophysics $\vert$ Harvard \& Smithsonian, Cambridge, MA 02138, USA}

\author{Duy N. Hoang}
\affiliation{Hamburger Sternwarte, University of Hamburg, Gojenbergsweg 112, 21029 Hamburg, Germany}

\author{Timothy W. Shimwell}
\affiliation{ASTRON, the Netherlands Institute for Radio Astronomy, Oude Hoogeveensedijk 4, 7991 PD Dwingeloo, The Netherlands}
\affiliation{Leiden Observatory, Leiden University, PO Box 9513, 2300 RA Leiden, The Netherlands}

\author{Paul Nulsen}
\affiliation{Center for Astrophysics $\vert$ Harvard \& Smithsonian, Cambridge, MA 02138, USA}
\affiliation{ICRAR, University of Western Australia, 35 Stirling Hwy, Crawley, WA 6009, Australia}

\author{William Forman}
\affiliation{Center for Astrophysics $\vert$ Harvard \& Smithsonian, Cambridge, MA 02138, USA}

\author[0000-0002-3984-4337]{Scott Randall}
\affiliation{Center for Astrophysics $\vert$ Harvard \& Smithsonian, Cambridge, MA 02138, USA}

\author{Yuanyuan Su}
\affiliation{University of Kentucky, 505 Rose street, Lexington, KY 40506, USA}

\author{Priyanka Chakraborty}
\affiliation{Center for Astrophysics $\vert$ Harvard \& Smithsonian, Cambridge, MA 02138, USA}

\author{Christine Jones}
\affiliation{Center for Astrophysics $\vert$ Harvard \& Smithsonian, Cambridge, MA 02138, USA}

\author[0000-0002-3031-2326]{Eric Miller}
\affiliation{Kavli Institute for 
Astrophysics and Space Research,
Massachusetts Institute of Technology, 70 
Vassar St, Cambridge, MA 02139}

\author{Mark Bautz}
\affiliation{Kavli Institute for 
Astrophysics and Space Research,
Massachusetts Institute of Technology, 70 
Vassar St, Cambridge, MA 02139}

\author[0000-0002-4737-1373]{Catherine E. Grant}
\affiliation{Kavli Institute for 
Astrophysics and Space Research,
Massachusetts Institute of Technology, 70 
Vassar St, Cambridge, MA 02139}

\begin{abstract}

We present a 368 ks deep Chandra 
observation of 
Abell~1240, a binary merging
galaxy cluster at a redshift 
of 0.195 with two 
 Brightest Cluster Galaxies (BCGs)
may have passed each other 
0.3 Gyr ago.
Building upon previous 
investigations involving GMRT, VLA, and 
LOFAR data, our study focuses on two
prominent extended radio relics at
the north-west (NW) and south-east (SE) 
of the cluster core.
By leveraging the
high-resolution  Chandra imaging, 
we have identified two distinct
surface brightness
edges 
at $\sim$ 1 Mpc
and 1.2 Mpc
NW and SE of
the cluster center,
respectively,
coinciding with the outer edges
of both relics.
Our 
temperature measurements
hint the edges to be shock front edges. 
The Mach numbers, derived from 
the gas density jumps, yield
$\cal{M}_{\rm SE}$ = 1.49$^{+0.22}_{-0.24}$ 
for the South Eastern shock
and $\cal{M}_{\rm NW}$ = 1.41$^{+0.17}_{-0.19}$
for the North Western shock.
Our estimated Mach numbers
are remarkably 
smaller compared to those derived
from radio observations
{($\cal{M}_{\rm SE}$ = 2.3 and
$\cal{M}_{\rm NW}$ = 2.4)}, 
highlighting the 
prevalence of 
a re-acceleration scenario over 
direct acceleration
of electrons from the 
thermal pool.
Furthermore, we compare the observed
temperature
profiles
across both shocks with that of predictions
from collisional vs. collisionless models.
Both shocks favor the Coulomb 
collisional model, but we could not 
rule out a purely collisionless
model due to
pre-shock temperature uncertainties.

\end{abstract}

\keywords{Galaxy cluster --- ICM --- Shock front --- Cosmology}
%TC:endignore

%%%%%%%%%%%%%%%%%%%%%%%%%%%%%%%%%%%%%%%%%%%%%%%%%%

%%%%%%%%%%%%%%%%% BODY OF PAPER %%%%%%%%%%%%%%%%%%

\section{Introduction}\label{sec:intro}

Galaxy clusters, the largest gravitationally
collapsed structures in the universe, play a 
vital role in understanding the evolution 
of cosmic structures. 
Over time, these clusters grow through
the complex processes of gas accretion 
from large-scale filaments and mergers of
smaller clusters and groups
{ \citep{2022ApJ...938...51A,2022arXiv221109827K,2023arXiv231002225Z}}. 
Such mergers are characterized by 
prodigious amounts of energy, 
long lifetimes 
extending over
billions of years, and 
vast 
physical scales
spanning several 
Mpc 
\citep{1999ApJ...521..526M,2022ApJ...935L..23S}. 
During these merger events, 
a substantial portion of the 
released gravitational energy is 
converted into thermal energy through
the generation of shocks and 
turbulence 
\citep{2007PhR...443....1M}.

However, beyond the thermal
component, a lesser fraction 
($<$ 1\%) of 
the shock energy might be channeled into 
the acceleration of cosmic rays (CRs)
\citep{2018MNRAS.478.2218H}. 
The presence of magnetic fields in galaxy 
clusters enables these 
accelerated CRs to 
emit synchrotron radiation,
which can be 
detected and studied 
using radio telescopes
\citep[e.g.,][]{1991SSRv...58..259J,1998A&A...332..395E,2010ApJ...721L..82C,2014IJMPD..2330007B,2019SSRv..215...16V}.
Radio relics, elongated and arc-like 
radio sources, have been identified as 
a prominent signature of cluster merger
shocks 
\citep[e.g.,][]{2009A&A...507..661B,2012A&ARv..20...54F,2014MNRAS.444.3130D,2016ApJ...818..204V,2022MNRAS.509.1160I}. 
Understanding the mechanism responsible 
for particle acceleration in these low-Mach
number cluster merger shocks is a 
significant challenge. 

Despite decades of observations
and simulations, the particle
acceleration  by the
merger shock is not yet fully
understood.
To explain the observed radio
relics, two main 
particle acceleration mechanisms 
have been proposed:
first, shock acceleration -- 
where particles 
{ from the thermal pool} 
gain energy 
through multiple crossings
of the shock front via 
Diffusive Shock Acceleration
(DSA; 
\citealt{1983RPPh...46..973D,2005ApJ...627..733M}).
According to the 
DSA theory, the efficiency 
of particle acceleration 
is expected to 
be very low for shocks with low
Mach number ($\cal{M}$ $<$ 4). 
Consequently, the existence of bright 
radio relics is puzzling within the 
context of standard DSA 
\citep[e.g.,][]{2005ApJ...627..733M,Kang_2012}.
On the other hand, re-acceleration
theory predicts shocks re-accelerate a 
population of 
pre-existing (``fossil") 
relativistic 
electrons via DSA 
\citep[e.g.,][]{2005ApJ...627..733M},
bypassing 
the low acceleration 
efficiency problem in shock
acceleration 
{ from the thermal pool}. 
Radio 
galaxies, commonly found 
in clusters, serve as 
good source candidates for 
these fossil electrons 
\citep[e.g.,][]{2017NatAs...1E...5V}.

X-ray and radio observations
of Abell 3411-3412 
provided
compelling evidence for particle 
re-acceleration at cluster 
merger shocks 
\citep{2017NatAs...1E...5V}. 
The discovery involves the 
observation of a tailed 
radio galaxy connected to a 
radio relic, 
accompanied by spectral 
flattening where the 
fossil plasma intersects
with the relic. 
Galaxy clusters with 
diametrically opposed double 
radio relics offer compelling 
platforms for investigating particle 
(re-)acceleration on Mpc scales
\citep{2011A&A...533A..35V,2018MNRAS.478.2218H}. 
The spatial arrangement of 
these double relics, 
emerging 
{ on the major
axis of an elongated
cluster}, 
points to head-on binary mergers 
of nearly equal-mass clusters 
occurring nearly in the plane 
of the sky 
\citep[e.g.,][]{2011MNRAS.418..230V,2017MNRAS.470.3465B}. 
This distinctive configuration 
minimises projection effects 
and offers a unique setting to 
explore particle (re-)acceleration 
without the complexities of 
relativistic electron mixtures 
along the line of sight (LOS) 
\citep{2013A&A...555A.110S}. 
Additionally, given the 
likelihood of mildly 
relativistic electron seed 
populations being tied to aged AGN 
outbursts, the presence of these 
double relics offers insights into the 
origin of relics -- whether they arise from 
direct acceleration of thermal pool 
electrons or from pre-existing fossil 
plasma within the 
{ intracluster medium (ICM)}
\citep{2018MNRAS.478.2218H,2022ApJ...927...80R}.

Abell 1240 (A1240 hereafter) is
a binary merging galaxy cluster
at redshift $z$ = 0.195
{ with a virial radius of 
1.9 $h_{70}^{-1}$ Mpc
\citep{2009A&A...503..357B}.
}
Observations in optical
band showed the galaxy clumps are 
separated
in the north-south direction by 
approximately 1.3 Mpc
and may 
have
passed each other 0.3 Gyr ago
\citep{2022ApJ...925...68C}.
A1240 is a well-studied cluster in
radio because of double radio
relics that are elongated over 
$\sim$
650 kpc (northern relic) 
and $\sim$ 1250 kpc 
(southern relic) 
in the 
east-west direction 
\citep{2009A&A...494..429B},
rendering it an excellent 
candidate for studying shock 
phenomena. 
Recent studies by 
\citet{2018MNRAS.478.2218H} 
presented spectral 
index maps derived from 
observations spanning the 
frequency range of 
145 MHz to 3 GHz 
(using LOFAR 
at 145 MHz, GMRT at 610 MHz, 
and JVLA at 2--4 GHz). 
They estimated shock Mach numbers
for both relics to be 2.4 $\pm$ 0.1 and
2.3 $\pm$ 0.1. 
Despite its potential, 
A1240 remains relatively 
unexplored in X-rays due to
the absence of deep 
observations from XMM-Newton or
Suzaku, and the previously 
existing Chandra observation
lacks the depth 
required for a comprehensive 
characterization of 
the cluster dynamics.
The primary goal of this
letter is to probe the ICM
properties and dynamics of two
shock fronts in A1240 using new deep
$\chandra$ observations. 
Table 
\ref{tab:obs_log} presents
all existing $\chandra$ 
observations 
of A1240, which were performed
with ACIS-I in the aim point. 

We adopted a cosmology of
$H_0$ = 67.8 km s$^{-1}$ Mpc$^{-1}$, 
$\Omega_{\Lambda}$ = 0.692, 
and $\Omega_{\rm M}$ = 0.308,
which gives a scale of 1$''$ = 3.346
kpc
at the redshift $z$ = 0.195 of A1240.
Unless otherwise stated, all reported error bars 
are at 68\% confidence level.

\begin{comment}
\begin{table}[ht!]
    \centering
    \caption{{\it Chandra} observation log}
    \begin{tabular}{ccccc}
\hline
Obs ID & Instrument & Exp. time & Obs Date & PI \\
 & & (ks) & &  \\    
\hline
4961 & ACIS-I & 51.4 & 2005-02-05 & Kempner \\
22646  & ACIS-I & 33.6 & 2020-02-09 & Andrade-Santos\\
22647 & ACIS-I & 23.8 & 2020-03-09 & Andrade-Santos \\   
22720  & ACIS-I & 20.5 & 2020-03-05  &        Kraft  \\
22965  & ACIS-I & 32.6 & 2020-02-22   &       Kraft \\  
23060 & ACIS-I & 19.8 & 2020-02-10 & Andrade-Santos \\
23061 & ACIS-I & 24.6 & 2020-02-27 & Andrade-Santos   \\
23145 & ACIS-I & 15.8 & 2020-02-09 & Andrade-Santos \\ 
23154 & ACIS-I & 21.8 & 2020-02-10 & Andrade-Santos  \\
23155 & ACIS-I & 16.8 & 2020-02-11 & Andrade-Santos \\ 
23165 & ACIS-I & 12.9 & 2020-02-22   &       Kraft \\ 
23176 & ACIS-I & 24.7 & 2020-03-02 & Andrade-Santos \\  
23180 & ACIS-I & 26.8 & 2020-03-08   &       Kraft  \\ 
23187 & ACIS-I  & 30.6 & 2021-01-08 & Andrade-Santos\\
\hline
    \end{tabular}
    \label{tab:obs_log}
\end{table}
\end{comment}

\begin{table}[ht!]
    \centering
    \caption{{\it Chandra} observation log}
    \begin{tabular}{cccc}
\hline
Obs ID & Filtered Exposure & Obs Date & PI \\
 & (ks) & &  \\    
\hline
4961  & 51.4 & 2005-02-05 & Kempner \\
22646 & 33.6 & 2020-02-09 & Andrade-Santos \\
22647 & 23.8 & 2020-03-09 & Andrade-Santos \\   
22720 & 20.5 & 2020-03-05 & Kraft \\
22965 & 32.6 & 2020-02-22 & Kraft \\  
23060 & 19.8 & 2020-02-10 & Andrade-Santos \\
23061 & 24.6 & 2020-02-27 & Andrade-Santos \\
23145 & 15.8 & 2020-02-09 & Andrade-Santos \\ 
23154 & 21.8 & 2020-02-10 & Andrade-Santos \\
23155 & 16.8 & 2020-02-11 & Andrade-Santos \\ 
23165 & 12.9 & 2020-02-22 & Kraft \\ 
23176 & 24.7 & 2020-03-02 & Andrade-Santos \\  
23180 & 26.8 & 2020-03-08 & Kraft \\ 
23187 & 30.6 & 2021-01-08 & Andrade-Santos \\
\hline
    \end{tabular}
    \label{tab:obs_log}
\end{table}

\section{Data preparation} \label{sec:data}
A1240 was observed with $\chandra$ during two 
epochs,
once in February 2005 for 51.4 ks
and later in February
2020 -- January 2021 for 316 ks divided 
into 13 observations. 
This yields a cumulative
exposure time of approximately 
368 ks (see Table \ref{tab:obs_log}
for detailed observation logs).
We conducted 
standard data reduction 
processes employing CIAO version 
4.15 and CALDB version 4.9.4 
distributed by the $\chandra$ X-ray 
Center (CXC). 
We have followed a standard data analyzing
thread
\footnote[4]{\url{http://cxc.harvard.edu/ciao/threads/index.html}}. 

All level 1 event files 
underwent reprocessing with 
the 
{\tt chandra$\_$repro} task by 
incorporating
the latest gain, charge transfer 
inefficiency correction, and 
filtering out the bad grades. 
VFAINT mode was used to improve 
the background screening. 
Flare-affected periods were 
detected and removed via the 
{\tt lc$\_$clean} script,
with resultant filtered exposure times 
documented in Table \ref{tab:obs_log}.
Employing 
the {\tt reproject$\_$obs} task,
all observations were repositioned 
to a shared tangent location before being 
combined.
The exposure maps in the 0.5-2.0 keV energy
bands were generated using 
{\tt flux$\_$obs}. 
%script by providing a weighting spectrum,
%which was generated using the {\tt %make$\_$instmap$\_$weights} 
%task with an absorbed APEC plasma emission model %and  
%a plasma temperature of 3 keV. 
To address underexposed detector edges, 
pixels with less than 15\% 
exposure relative to the combined 
duration were zeroed. 

Point sources were identified using {\tt 
wavdetect} with a range of
wavelet radii between  1--16 pixels.
The detection threshold 
was set to $\sim$ 10$^{-6}$, 
ensuring fewer than one spurious 
source detection per CCD.
We used blanksky background observations 
to model the non-X-ray 
background, emission from foreground structures 
(e.g., Galactic Halo and Local Hot Bubble) 
along the observed direction
and unresolved faint background sources. 
The blanksky background files were 
generated using the {\tt blanksky} task 
and then 
reprojected to match the coordinates 
of the observations. 
We finally tailored the resulting 
blanksky background 
to match the 
9.5--12 keV count rates in
our observations.

\begin{figure*}
    \centering
    \includegraphics[width=0.9\textwidth]{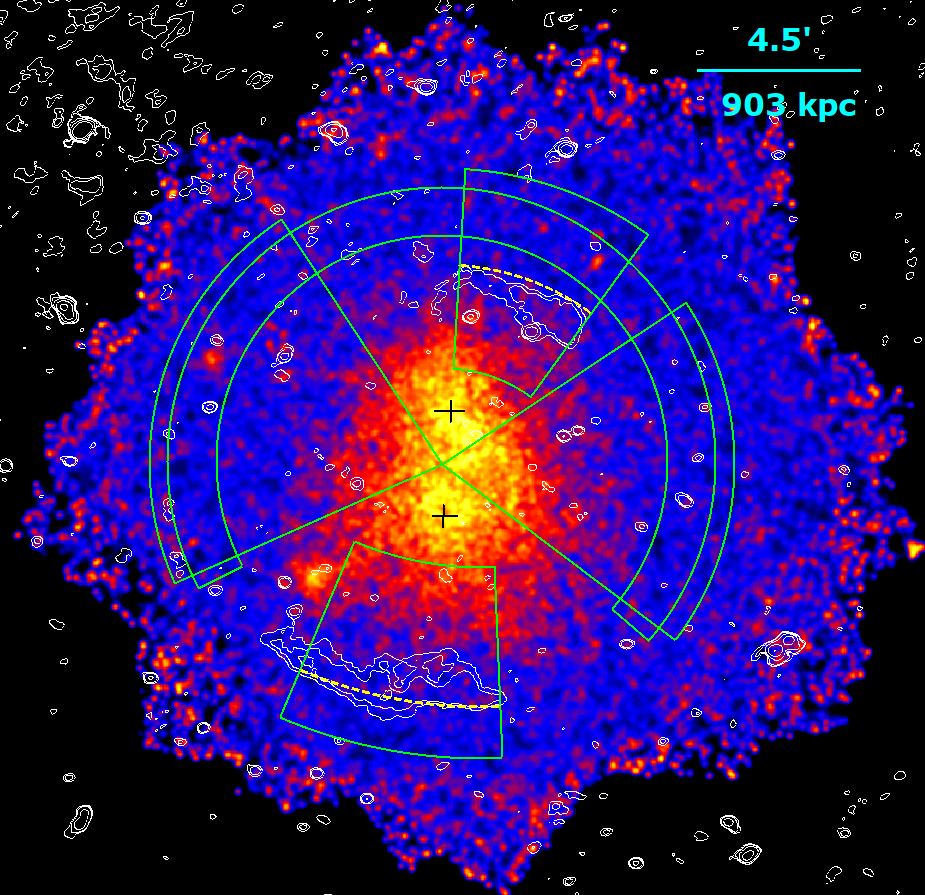}
    \caption{Exposure corrected blanksky-background
    subtracted $\chandra$ image of A1240
    in the 0.5--2 keV energy band.
    Both BCGs are shown in black.
    White contours show LOFAR 143 MHz radio 
    contours of [-3, 3, 6, 12, 24, 48]
    $\times$ $\sigma_{\rm rms}$, where 
    $\sigma_{\rm rms}$ = 165 $\mu$Jy/beam, taken from
    \citet{2018MNRAS.478.2218H}.
    Green sectors in NW, SE, and W directions
    are used for extracting surface brightness
    and spectral analysis. 
    The yellow dashed
    curves indicate the positions of shock
    fronts. The large green annulus running from
    W, N to E represents the region used for
    pre-shock temperature measurement.}
    \label{fig:image}
\end{figure*}

\section{Results}{\label{sec:results}}
We present deep 
$\chandra$ observations of 
A1240 to investigate 
the merger shocks and associated radio 
relics in the cluster.

\subsection{Imaging analysis}{\label{sec:imaging_analysis}}
Figure \ref{fig:image}
displays the $\chandra$ image of
A1240 in the 0.5 -- 2 keV energy band, 
revealing the X-ray emission of the 
cluster superimposed onto radio contours.
The existing
shallow X-ray emission exhibits an 
elongated structure in the north-south
direction, 
indicating a double X-ray morphology, 
consistent with a slightly asymmetric 
merger
\citet{2018MNRAS.478.2218H}.
Radio observations from the
VLA (325, 1400 MHz)
have indicated that A1240 experienced 
a major merger between
two subcluster 
cores, which passed each other in the 
north-south direction, 
resulting in two merger shocks 
currently ahead of each subcluster
\citep{2009A&A...494..429B}. 
These merger shocks are 
responsible for (re)accelerating
electron populations, 
giving rise to extended radio 
relics that shine brightly in the radio band 
%\textcolor{red}{it sounds as if Maxim had 
%presented this finding for A1240}. 
The northern and southern radio relics
are located at $\sim$ 900 kpc and 1.1 Mpc from
the cluster X-ray center
\citep{2018MNRAS.478.2218H}. Both relics
are extended in the EW direction
($\sim$ 0.7 -- 1.3 Mpc wide;
as seen in Figure \ref{fig:image}),
perpendicular to the main axis of 
the X-ray emission, similar to the
double relics observed in 
several other galaxy clusters
\citep[e.g.,][]{2011A&A...533A..35V}.

\begin{figure*}
    \centering
    \begin{tabular}{cc}
            \includegraphics[width=0.5\textwidth]{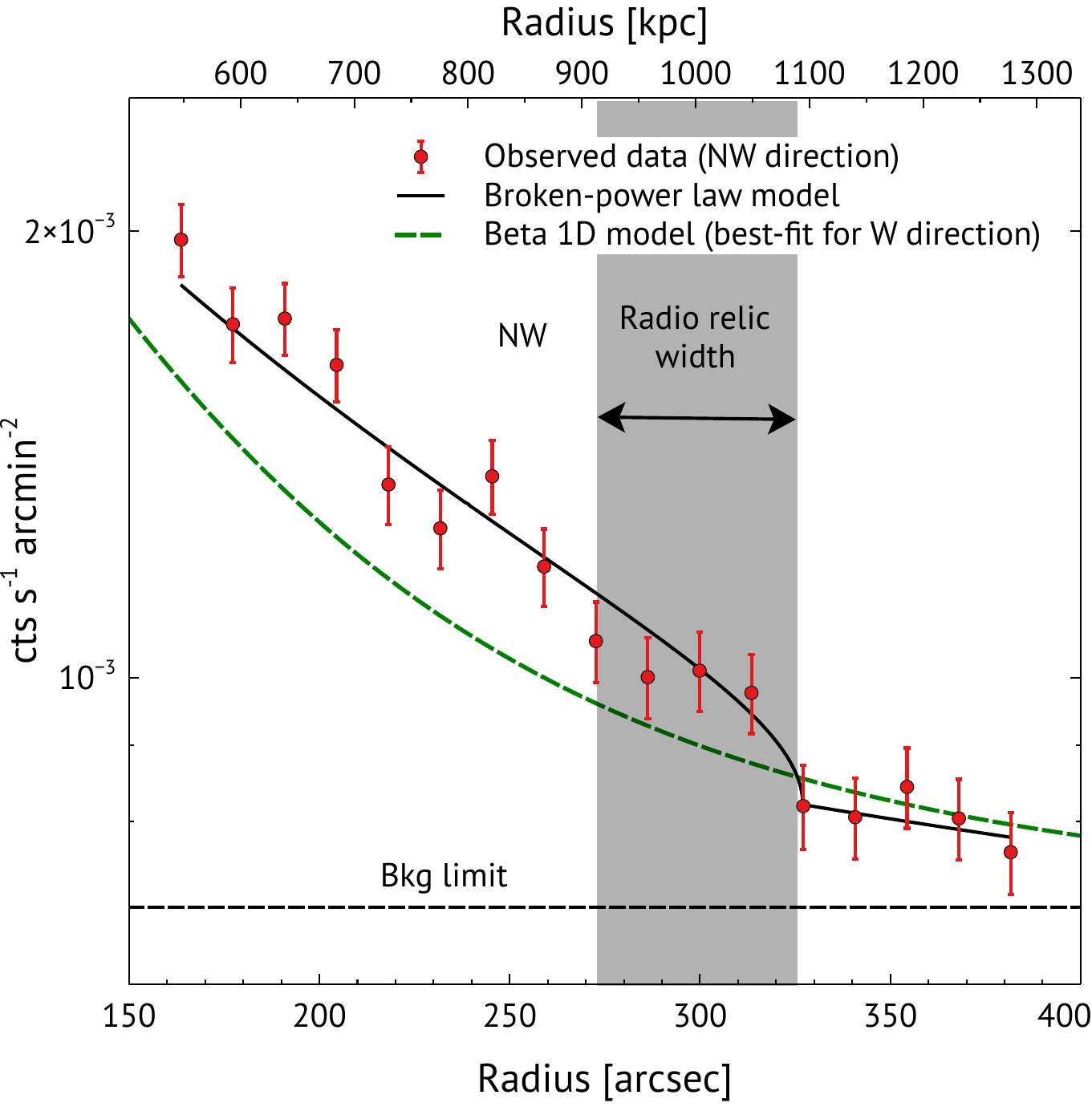} &  \includegraphics[width=0.5\textwidth]{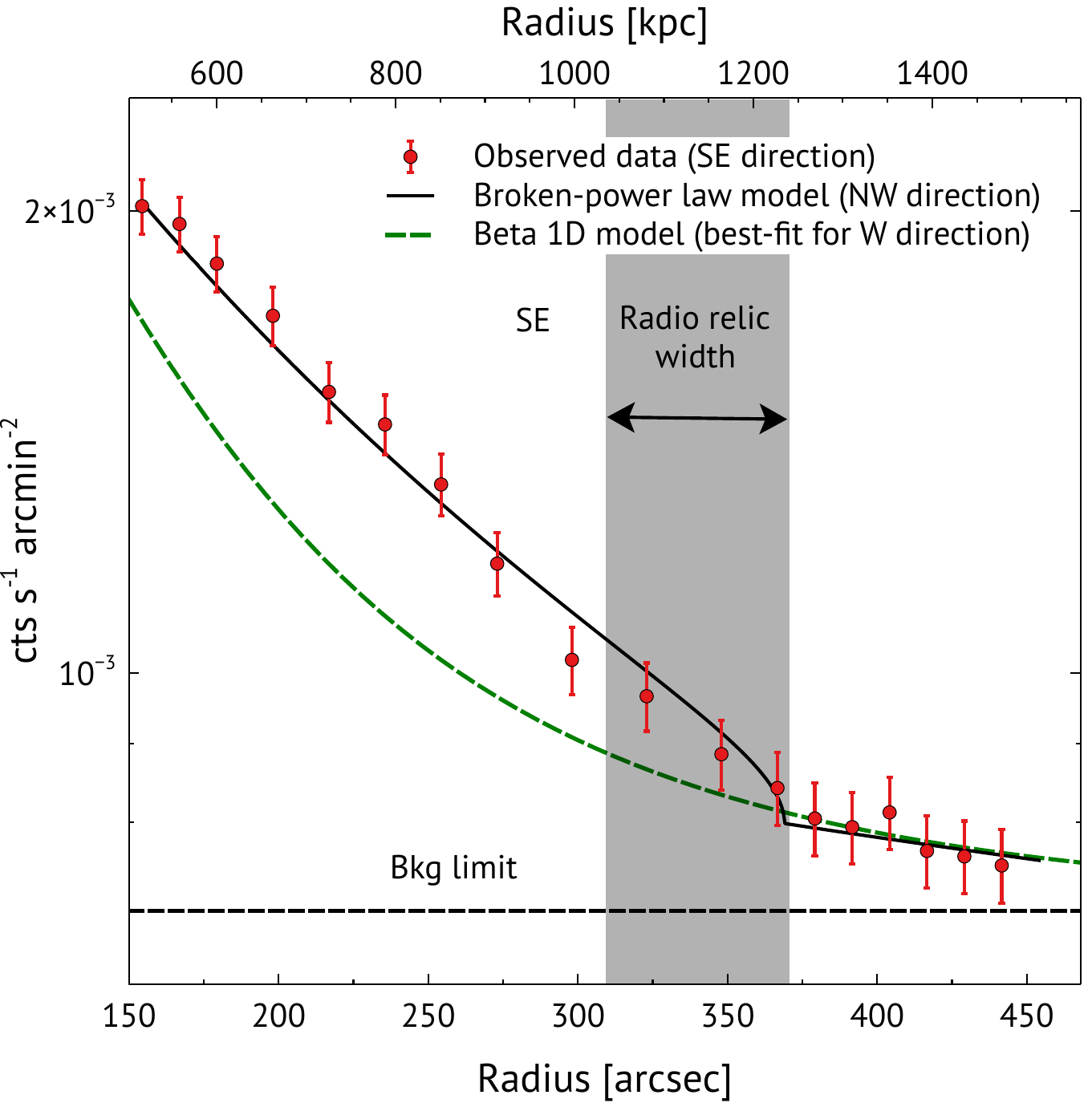}\\
    \end{tabular}
    \begin{tabular}{c}
 \includegraphics[width=0.5\textwidth]{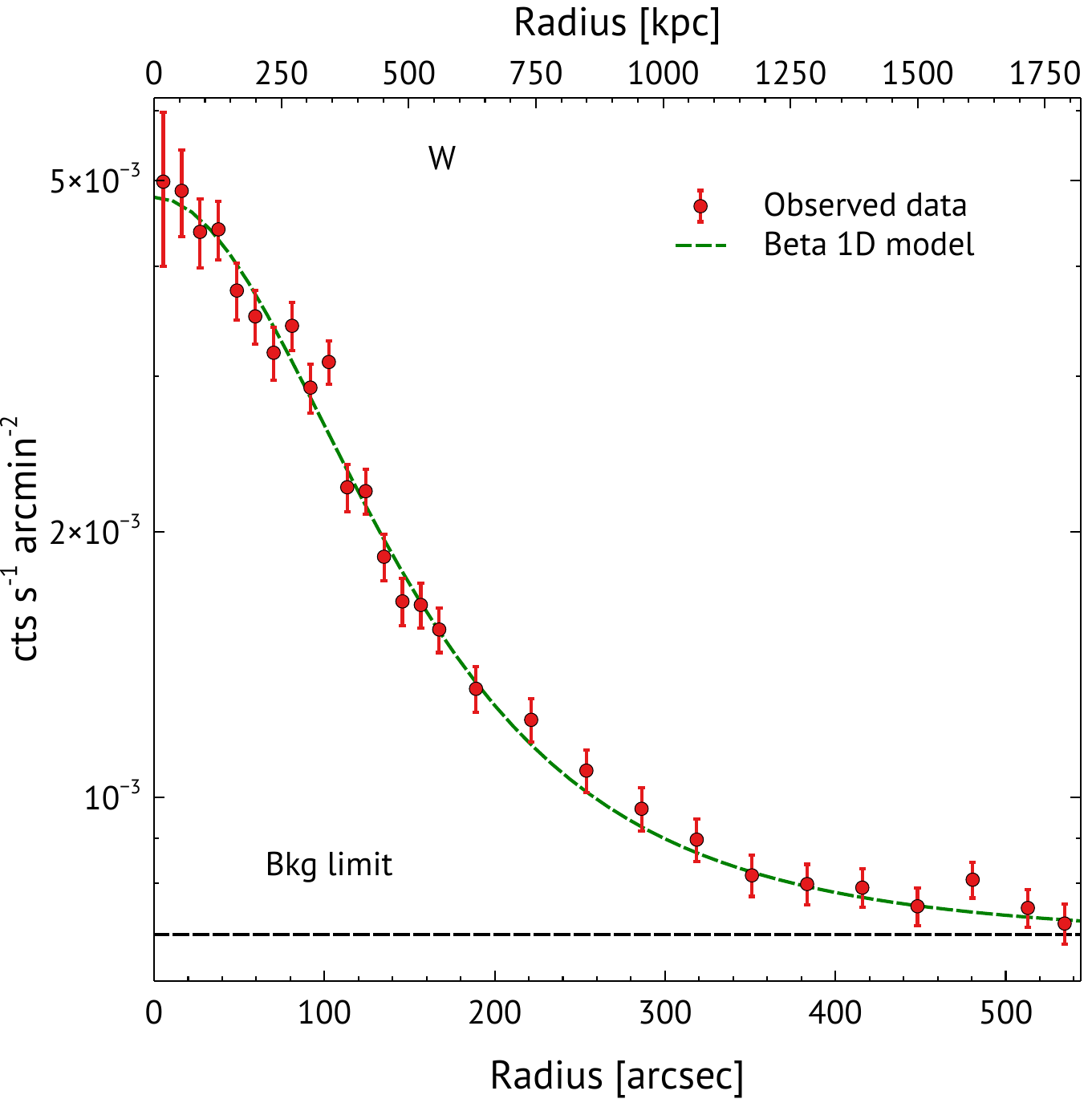}\\
    \end{tabular}
    \caption{ {\it Top-Left}: Surface
    brightness profile across the 
    NW relic (red). Black curve shows 
    the best-fit broken powerlaw model.
    Best-fit electron 
    density is shown in inset.
    Green dashed curve
    shows best-fit beta model
    for the surface brightness
    profile in the W direction. 
    It is shown here for 
    comparison. 
    The black horizontal dashed line shows the
    sky + instrument background limit.
    Grey shaded region represents
    the width of the relic. 
    {\it Top-Right}: similar to {\it Top-Left} 
    but for SE relic.
    {\it Bottom}: Surface brightness profile
    of A1240 extracted from the sector
    in the W direction. The green
    dashed curve
    shows the best-fit 1D-$\beta$ model, same is shown in 
    {\it Top-Left} and 
    {\it Top-Right}
    panels.}
    \label{fig:sur_bri}
\end{figure*}

Previous $\chandra$ observations of 
A1240 were too shallow (52 ks), 
hence low signal-to-noise, to properly 
pinpoint the apparent location of both 
shock front edges, which is 
crucial to derive Mach numbers of both shocks
\citep{2018MNRAS.478.2218H}. With new 
deep $\chandra$ observations, 
we examine both shock front edges 
by extracting surface brightness 
profiles across both edges. 
Figure \ref{fig:sur_bri} shows the 
resulting radial
surface brightness profiles as a 
function of distance from the A1240 core
in the 0.5 -- 2keV energy band. 
The extracted profiles across both shock 
front edges exhibit shapes consistent 
with the expected projection of a 
3D density discontinuity 
\citep[e.g.,][]{2000ApJ...541..542M,2023ApJ...944..132S,2023arXiv230804367W}.
To further quantify the surface brightness 
edges, we fit the profiles with a 
spherically symmetric
discontinuous double power-law
model projected along the line of sight ($l$),
\begin{equation}
   S_X \propto \int{n_{\rm e}^2\  dl},
\end{equation}
where,
\begin{equation}
    n_{\rm e}(r) \propto
    \begin{cases}
    \left(\frac{r}{r_{\rm edge}}\right)^{-\alpha_{\rm post}}, & \text{if $r < r_{\rm edge}$}\\
    \frac{1}{C} \left(\frac{r}{r_{\rm edge}}\right)^{-\alpha_{\rm pre}}, & \text{if $r \geq r_{\rm edge}$}
    \end{cases}
\end{equation}
where $n_e(r)$ is the 3D electron density at a radius $r$, 
$r_{edge}$
is the radius of the putative edge, 
$C$ is the 
density jump, 
and $\alpha_{\rm post}$ and $\alpha_{\rm pre}$
are the slopes inside and 
outside the edge, respectively. 
Additionally, a constant term was 
introduced to the model to account for 
residual background after blank-sky subtraction, 
and its best-fit value was consistent with zero, 
indicating successful background elimination.

By projecting the estimated emission 
measure profiles onto the sky 
plane and using least-square fitting, we fit
the observed surface brightness profiles,
varying $\alpha_1$, $\alpha_2$, $r_{\rm edge}$, 
and the 
$C$ as free parameters.
The best-fit surface brightness profiles
are shown in Figure \ref{fig:sur_bri}. 
The best-fit slopes for the SE edge
are 
$\alpha_{\rm post}$ = 2.25 $\pm$ 0.2
$\alpha_{\rm pre}$ = 0.8 $\pm$ 0.1, and 
for the
NW edge are $\alpha_{\rm post}$ = 1.15 $\pm$ 0.3,
$\alpha_{\rm pre}$ = 0.7 $\pm$ 0.3.
The electron density jumps by a factor of
$\rho_{\rm post}$/$\rho_{\rm pre}$ = 1.7 $\pm$ 0.3
($\chi^2$/dof = 9.2/12)
across the SE edge and 
$\rho_{\rm post}$/$\rho_{\rm pre}$ = 1.6 $\pm$ 0.2
($\chi^2$/dof = 13/12), 
across the NW edge.
Assuming these edges represent shock fronts, 
the derived density jump factors
correspond to Mach number of 
$\cal{M}_{\rm SE}$ = 1.49$^{+0.22}_{-0.24}$ (SE edge)
and 
$\cal{M}_{\rm NW}$ = 1.41$^{+0.17}_{-0.19}$
(NW edge),
estimated from the Rankine-Hugoniot 
jump condition, defined as
\begin{equation}\label{eq:mach}
    {\cal{M}} = \left [\frac{2 \ C}{\gamma + 1 - C\left (\gamma - 1 \right)}\right]^{\frac{1}{2}},
\end{equation}
where $C$ = $\rho_{\rm post}/\rho_{\rm pre}$ 
and for a monoatomic gas 
$\gamma$ = 5/3.
We find the
location for the SE edge
is 1230 $\pm$ 50 kpc and 
1095 $\pm$ 33 kpc for 
the NW edge from the
cluster center.
The uncertainties on model parameters
are estimated by varying other model parameters
freely.

We also extract surface brightness profile
along the
W direction (from the
sector shown in Figure \ref{fig:image}) 
to compare with that
of SE and NW directions. 
Figure \ref{fig:sur_bri} (bottom) shows the
resulting surface brightness profile in the
W direction. We fit this profile
with a 1D-$\beta$ model, as shown in 
Figure \ref{fig:sur_bri}.
In 
contrast to the NW and SE directions,
no surface brightness discontinuity is 
found in the W direction, which
confirms
the 
north-south axis as the merging 
direction and potential
locations of the shock fronts.

\subsection{Spectral analysis}{\label{sec:spectral_analysis}}
To measure the temperature 
across both surface brightness edges, 
we divided the SE and NW sectors 
(same used for surface birghtness
profiles)
into four regions for each sector.
We did the same for the E and W sectors
to compare 
temperatures with that of
the 
temperature across the shock.
Each region was carefully chosen 
to ensure a minimum of $\sim$
2000 background-subtracted counts 
in the 0.6 -- 7.0 keV energy band. 
This lower limit was set to guarantee 
a sufficient number of counts for 
accurate temperature measurements and
to achieve uncertainties within $\sim$
30\% in the faint pre-shock regions at 
a 68\% confidence level.

For each selected region,
we extracted spectra from individual 
observations and grouped them to 
contain a minimum of 30 counts per 
spectral channel. 
Background spectra were also 
extracted from the blank-sky event 
files and subtracted from the source 
spectra before fitting 
\citep{2016ApJ...820L..20D,2021MNRAS.501.3767S}.
The extracted spectra from each region
were fitted 
in XSPEC with an absorbed 
single-temperature thermal emission model, 
{\tt PHABS $\times$ APEC} 
\citep{Smith_2001}. 
In the spectral fitting process, 
we fix the redshift 
to z=0.195, and the absorption was set to the 
Galactic value of $N_{\rm H}$ = 
2.65 $\times$ 10$^{20}$ cm$^{-2}$, following 
\citet{2005A&A...440..775K}. 
The best-fit 
parameters were obtained by reducing 
C-statistics \citep{1979ApJ...228..939C}, 
which provides robust estimates 
even for 
spectra with low counts
{ \citep{2023arXiv231212712C}}.
To improve the stability of the fitting process, 
we fixed the 
metallicity to an typical
value of 0.3
{ \citep{2017MNRAS.470.4583U,2022MNRAS.516.3068S,2023arXiv231004499M}.}
This choice 
was made since leaving the metallicity as a free 
parameter led to poorly constrained results, as 
observed in previous studies by 
\citet{2012MNRAS.423..236R} and
\citet{2023ApJ...944..132S}. 
We adopted the solar abundance table of 
\citet{2009ARA&A..47..481A}.

\begin{figure*}
    \centering
    \begin{tabular}{cc}
\includegraphics[width=0.485\textwidth]{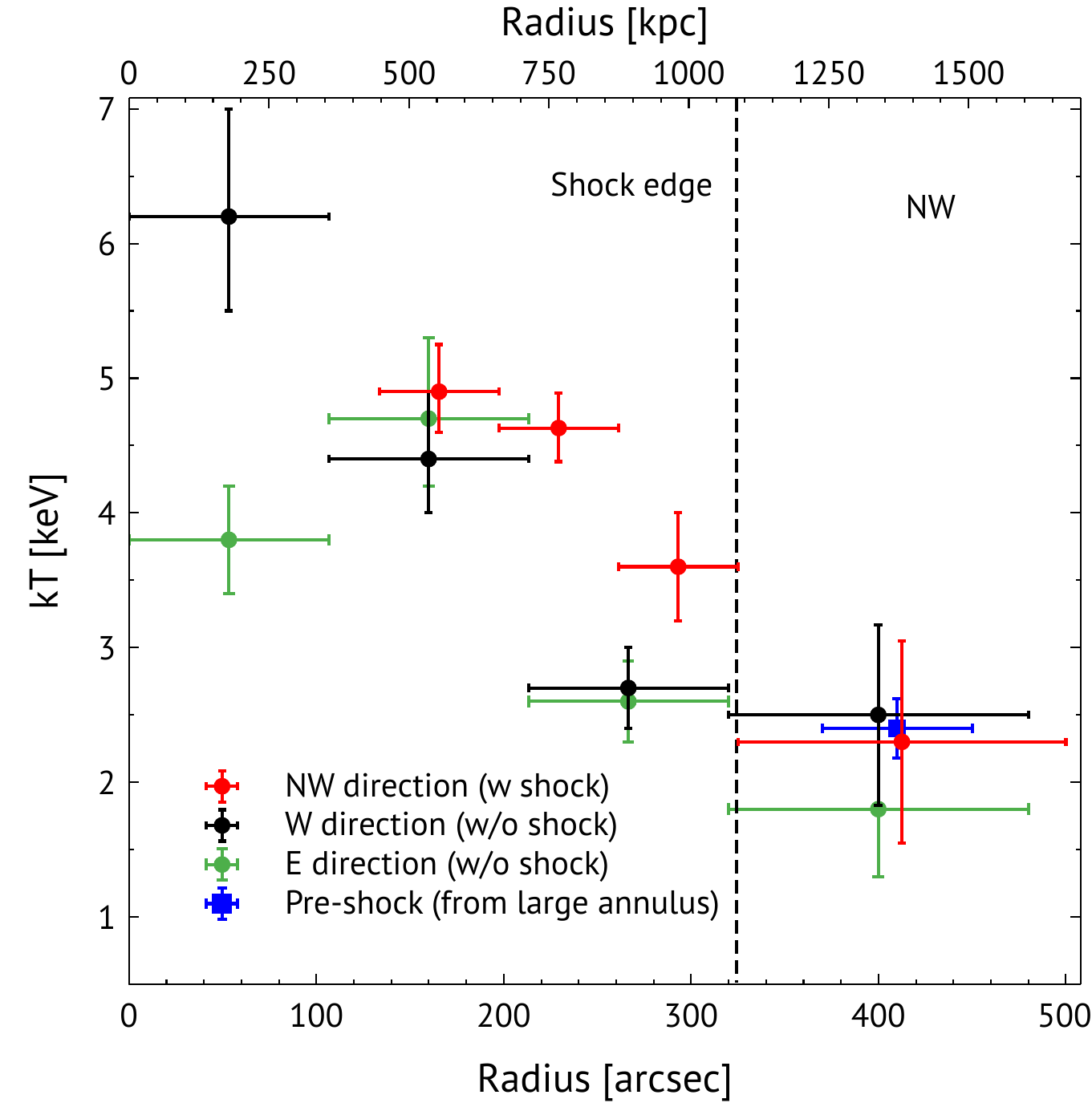} &  \includegraphics[width=0.485\textwidth]{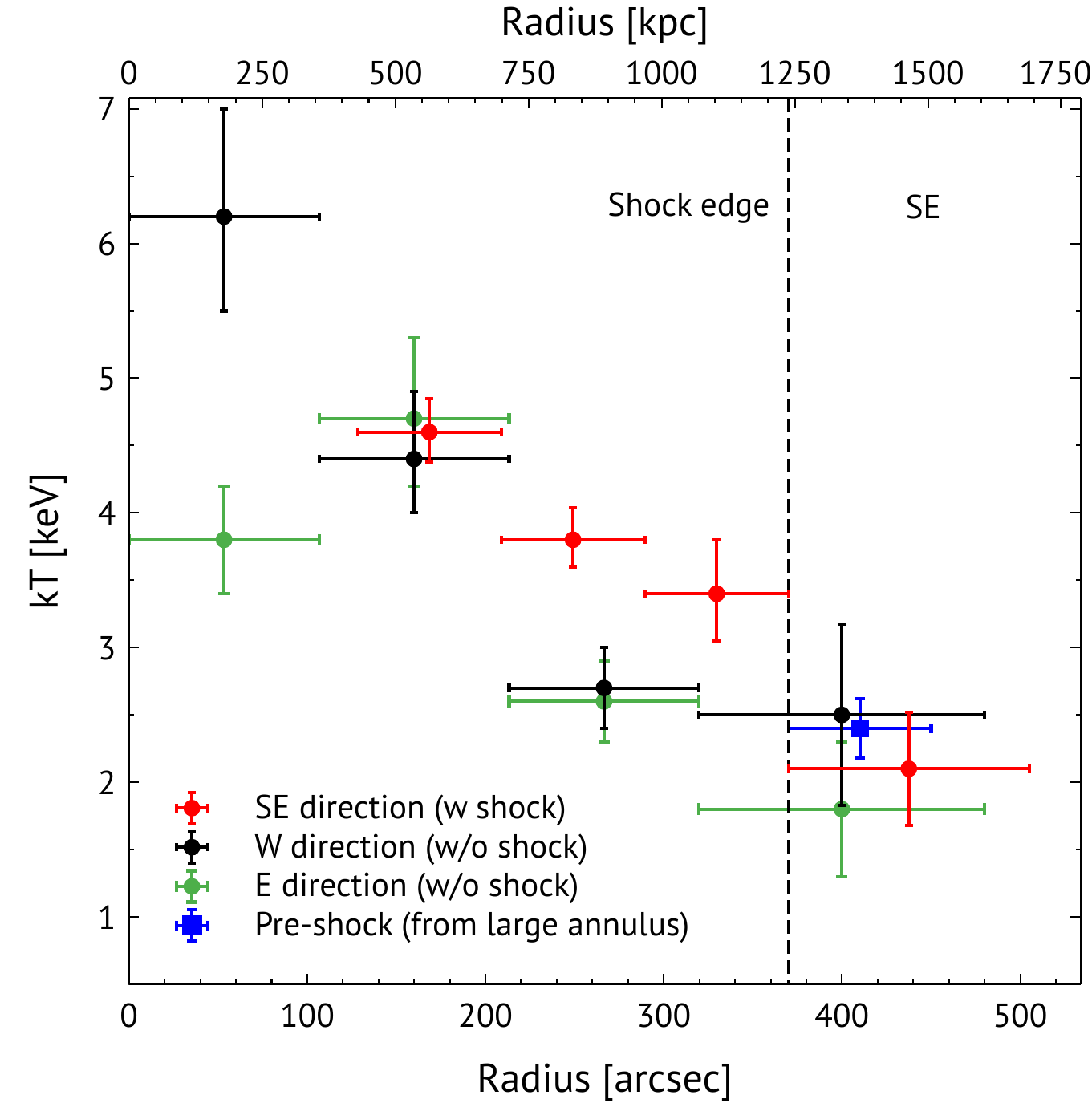}\\
    \end{tabular}
    \caption{{\it Top-Left}: Temperature
    profile across the 
    NW relic (red). Temperature
    profile along W direction (black)
    and 
    E (green) direction
    are plotted for comparison.
    {\it Top-Right}: Temperature
    profile across the 
    SE relic (red).
    The vertical dashed line represents
    the location of the shock from
    the cluster center.
    In all three figures blue
    represents the temperature
    measurement from a large annulus 
around the cluster, as shown
in Figure \ref{fig:image}.}
    \label{fig:temp_profile}
\end{figure*}

Figure \ref{fig:temp_profile} illustrates 
the best-fit projected temperature profiles 
in the different directions.
We measure temperature across the surface
brightness edges in 
the NW and SE directions.
It is evident that the temperature 
decreases from 
the cluster center.
To isolate the temperature jump
attributed to the shock, 
we also measure
temperatures in the
W and E directions and compare
them with NW and SE directions.
In the NW, SE, and W directions, 
we observe that the temperature decreases 
from the cluster center outwards,
while the temperature 
increases from the cluster center
in the E direction. 
Temperature measurements across all
radial bins are found to be consistent 
with each other, except for the third 
radial bin in the NW and SE directions,
where we measure $\sim$ 2$\sigma$ hotter gas
compared to the 
other two directions.

{ 
Given the large uncertainties
associated with the temperatures from 
the last radial bins in the SE and NW 
directions, we opted for a 
broader approach to measure the pre-shock
temperature with smaller uncertainty.
We measure the pre-shock 
temperature from a large circular
annulus region around
the cluster spanning a 
radial range 
of 1.25–-1.5 Mpc, 
as illustrated in Figure \ref{fig:image}.
The convergence of surface brightness 
profiles in the NW, SE, and W directions 
at $\geq$ 1.25 Mpc from the cluster center, 
as seen in Figure \ref{fig:sur_bri}, 
supports this choice. 
Moreover, gas temperatures in the
NW, SE, W, and E directions at
$\geq$ 1.25 Mpc exhibit consistency 
within their 1$\sigma$ uncertainties.
These observations collectively 
indicate uniform ICM properties
of A1240 at a 
radius $\geq$ 1.25 Mpc in all directions, 
justifying the selection of a circular
annulus around the cluster for 
pre-shock temperature measurement.
We measure a pre-shock temperature
of $kT_{\rm pre}$ = 2.38 $\pm$ 0.21
with a $\sim$ 8\% temperature uncertainty 
(1$\sigma$ level),
which is 2.5 and 3.4 times smaller compared
to the pre-shock temperature 
measurement uncertainties 
along NW and SE direction. 
}

{ The temperature 
jump detected for the NW 
direction is 
$\sim$ 1.6 at a 2.5$\sigma$
significance level
and for the SE direction
is $\sim$ 1.7 
at a 2.7$\sigma$ significance
level.
The radial distances
of both temperature jumps
from the cluster center
coincide with 
the locations of 
corresponding 
surface brightness 
edges.
We note that
the temperature jump has been measured
with respect to the pre-shock temperature
measured from the large circular
annulus 
around the cluster.}
These measurements hint
the presence
of shock heated gas in
the NW 
and SE directions.

{ For a shock discontinuity,
Rankine--Hugoniot jump conditions
relates electron density jump
and temperature jump as
\citep{1959flme.book.....L}
\begin{equation}\label{eq:temp_jump}
    t = \frac{\zeta - C^{-1}}{\zeta - C},
\end{equation}
where $t$ = $T_{post}$/$T_{pre}$
is the temperature jump
across
the shock front,
and $\zeta$ = ($\gamma$+1)/($\gamma$$-$1).
We derive the Mach numbers 
associated
with the 
above observed temperature 
jumps by 
using Equation \ref{eq:mach}
and \ref{eq:temp_jump}.
}
For the SE edge, the estimated 
Mach number is
$\cal{M}_{\rm SE}$ =
1.71 $\pm$ 0.25, while for the NW edge, 
$\cal{M}_{\rm NW}$ = 1.57 $\pm$ 0.34.
%{ We note given that the cluster is
%in later stages of a major merger,
%our choice of using pre-shock temperature
%measured from a large annulus 
%around the cluster might
%add some uncertainity 
%in Mach number estimation due to
%the distortions/perturbations from 
%the merger.}
These Mach numbers are 
consistent with those
obtained from surface brightness
profiles ($\cal{M}_{\rm SE}$ = 1.49$^{+0.22}_{-0.24}$ 
and $\cal{M}_{\rm NW}$ = 
1.41$^{+0.17}_{-0.19}$).
{ For comparison, we
estimate the Mach numbers by 
considering the temperature jump 
relative to the last radial bins 
of the temperature profiles in
both NW and SE directions.  
For SE edge, the
estimated Mach number is 
$\cal{M}_{\rm SE}$ = 1.62 $\pm$ 0.41
and for NW edge, $\cal{M}_{\rm NW}$ 
= 1.56 $\pm$ 0.52.}

\section{Electron heating mechanism}
Shock fronts in galaxy clusters play 
a pivotal role in redistributing energy
and accelerating particles, leading to 
a variety of observable phenomena.
Thus, understanding the thermal behavior
behind these shock fronts is crucial for 
deciphering the underlying
physical processes. 
In this section, we compare 
the application of two fundamental
models, the Coulomb collisional 
model and the instant-equilibrium model, 
to investigate the thermodynamics 
behind a shock front
\citep{2005ApJ...627..733M,2012MNRAS.423..236R,2022MNRAS.514.1477R}.

The Coulomb collisional model, 
built upon collisional ionization equilibrium 
assumptions, considers the time-dependent 
interactions between charged particles in the 
post-shock region. 
The
Coulomb collisional model predicts when 
a shock propagates through collisional
plasma it heats heavier ions dissipatively 
within a narrow region of a few ion-ion 
collisional mean free paths
\citep{1962pfig.book.....S,1988xrec.book.....S,1998MNRAS.293L..33E,2007PhR...443....1M}.
{
The thermal equilibrium time
of the ions among themselves
is roughly $\sqrt{(m_{p}/m_e)}$
= 43 times faster than the 
thermal equilibration time
between the electrons and the ions.
Thus, the ion kinetic energy
that is thermalized
by the 
shock is shared among the ions 
much more quickly than
it is shared with the electrons.
}
%In contrast, electrons, having
%significantly higher thermal 
%velocity than the shock,
%experience a less pronounced 
%temperature increase compared to %ions. 
Initially, electrons are adiabatically 
compressed in shocks and
later equilibrate with 
ions through Coulomb scattering,
where the equilibration time-scale is governed by
\citep{1962pfig.book.....S,1988xrec.book.....S},
\begin{equation} \label{eq:teq}
    {t_{\rm eq}}(e,p) \approx 6.2 \times 10^8 ~{\rm yr}  \left( \frac{{T}_{\rm e}}{10^8 \rm K} \right)^{3/2} \left( \frac{{\rm n}_{\rm e}}{10^{-3} {\rm cm}^{-3}} \right)^{-1}
\end{equation}
where $T_{\rm e}$ and $n_{\rm e}$ are the
electron temperature and density, respectively.
Electron temperature 
rises at the shock front 
via adiabatic compression,
\begin{equation}\label{eq:Te} 
        {T}_{\rm e,2} = {T}_{\rm e,1} \left( \frac{\rho_{\rm post}}{\rho_{\rm pre}} \right)^{\gamma -1}
\end{equation}
where $\rho_{\rm pre}$ and $\rho_{\rm post}$ are the gas density
in the pre-shock and post-shock regions. 
Electron and ion temperatures then 
subsequently equilibrate via Coulomb collision at a rate
given by, 
\begin{equation}\label{eq:rate}
        \frac{{\rm d}T_{\rm e}}{{\rm d}t} = \frac{{T_{\rm i}} - {T_{\rm e}}}{ t_{\rm eq}} 
\end{equation}
where $T_{\rm i}$ is the ion temperature.
Since the total thermal
energy density is conserved, 
the local mean gas temperature, $T_{\rm gas}$ is constant with time, 
where $T_{\rm gas}$ is given by,
\begin{equation}\label{eq:Tgas}
        { T_{\rm gas}} = \frac{n_{\rm e}T_{\rm e}+n_{\rm i}T_{\rm i}}{n_{\rm e}+n_{\rm i}} = \frac{ 1.1T_{\rm e} + T_{\rm i}}{2.1}
\end{equation}
where $n_{\rm i}$ is the ion density, with $n_{\rm e} = 1.1 n_{\rm i}$.

Alternatively, instant 
equilibration model
predicts electrons are strongly heated at
the shock front, similar to heavier ions, 
via magnetic fields, hence collisionless heating.
The equilibration timescale in the instant heating model
is much shorter than $t_{\rm eq}$ and
post-shock electron temperature is determined
by the Rankine-Hugoniot jump conditions 
\citep{2007PhR...443....1M}.

To compare both model predictions with the
observed temperature profile, we estimate
post-shock electron temperature from both
models using 
equations \ref{eq:teq}--\ref{eq:Tgas}
and projected them along the line of sight using
\citep{1998MNRAS.293L..33E},
\begin{equation} \label{eq:project}
    \langle { T} \rangle =\int_{b^2}^{\infty}\frac{\epsilon( r){ T_{\rm e}( r) dr^2}}{\sqrt{r^2 - b^2}} \bigg/ {\int^{\infty}_{b^2}\frac{\epsilon( r){ dr^2}}{\sqrt{r^2 - b^2}}}
\end{equation}
where $\epsilon(r)$ 
is the emissivity at physical radius $r$
and $b$ is the distance from the shock front projected onto
the sky plane.

\begin{figure*}
    \centering
    \begin{tabular}{cc}
\includegraphics[width=0.485\textwidth]{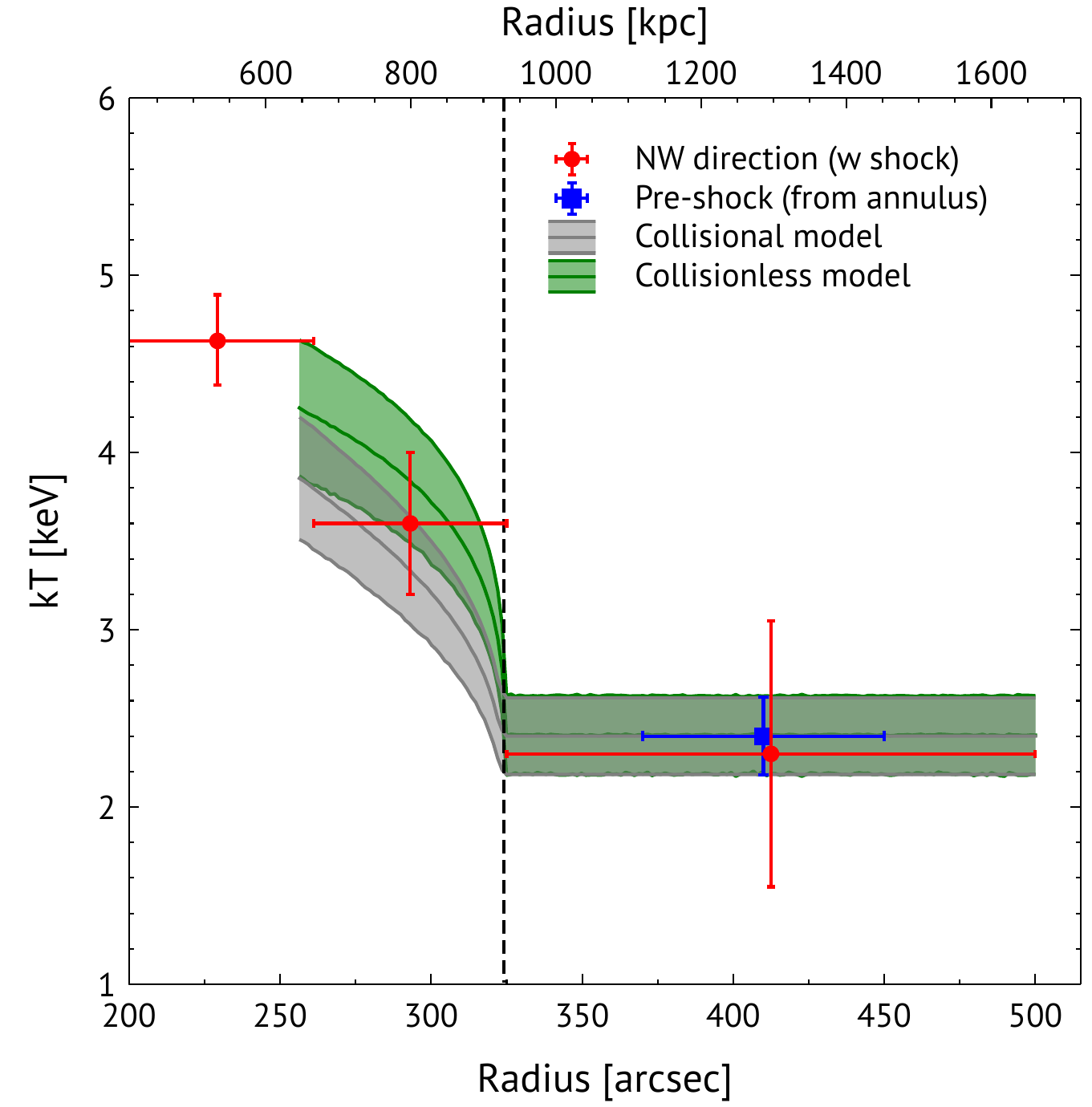} &  \includegraphics[width=0.485\textwidth]{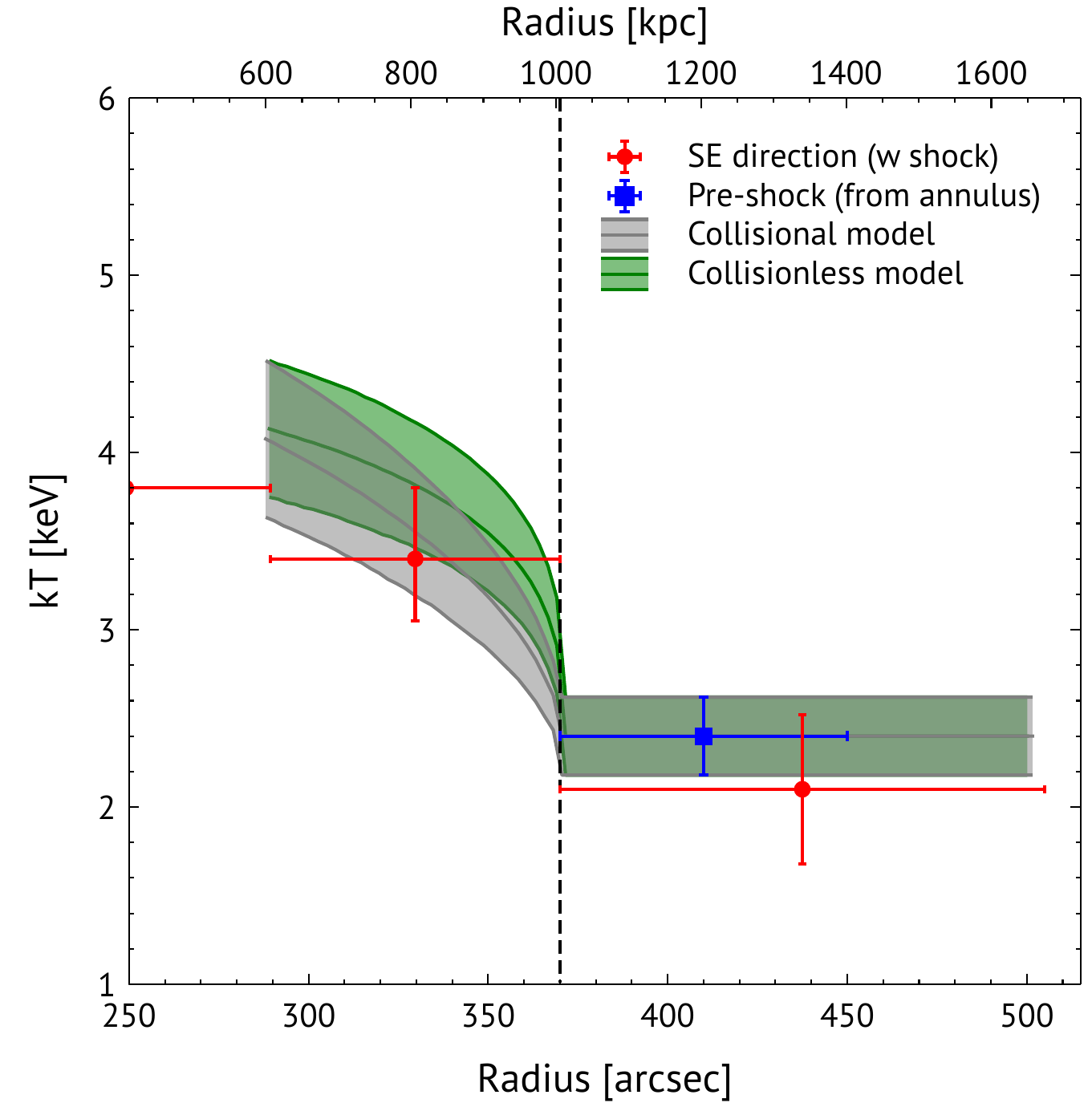}\\
    \end{tabular}
    \caption{
    Comparison of the temperature profiles across both shock
fronts with the predicted electron temperature profiles based on
instant collisionless model (green) and Coulomb collisional model (gray).
    {\it Left}: across the NW relic,
    {\it Right}:  across the SE relic.
    { Both models are only 
    valid for pre and post-shock temperatures
    only since the model estimations do not
    include the inherent temperature
    gradient of the cluster.}}
    \label{fig:colli_vs_ins}
\end{figure*}

Equation \ref{eq:project} gives 
the emission-weighted
temperature, which is the ideal 
spectroscopic-like
temperature if measured with a perfect instrument
with flat energy response. 
{In reality, X-ray telescopes
do not have flat response.
We, therefore, convolve 
the outcomes of Equation 
\ref{eq:project} for}
both models
with $\chandra$'s actual response file to predict
what we expect to measure. 
For detailed analysis,
we refer
readers to \citet{2022ApJ...935L..23S}.
Figure \ref{fig:colli_vs_ins}
shows the comparison between
model predictions and observed temperature
profiles of both shocks. 
Since the pre-shock temperature is the primary 
source of
uncertainty in this measurement, 
we predominantly use temperature
measurement from a large annuli around the
cluster as the pre-shock temperature.
For both shock fronts, our measurements
favor the collisional model over the
collisionless model, similar to
A2146 \citep{2022MNRAS.514.1477R}.
However, we can 
not rule out the purely
collisionless model due
to the large uncertainties in 
ICM temperature measurements.
{ This prevents us from 
definitively favoring 
one model over the other.
}

\section{Particle acceleration mechanism}
Shocks can (re-)accelerate
some ICM electrons to relativistic energies, 
causing them to emit detectable radio 
emission in the presence of the cluster's 
large-scale 
$\mu$G magnetic field 
\citep{1987PhR...154....1B,2012A&ARv..20...54F,2014IJMPD..2330007B}.
Merger shocks are often characterized
by low
Mach numbers, with typical 
values 
{ ranging 1.5--3}, as inferred from X-ray observations
\citep{2013PASJ...65...16A}. 
Despite the growing understanding of these 
processes, the precise efficiency of electron 
acceleration by low Mach number shocks 
associated 
with the
radio relics remains an intriguing puzzle
\citep[e.g.,][]{2017NatAs...1E...5V,2020A&A...634A..64B}.
We discuss the implications of our 
findings in the 
context of shock acceleration scenarios, 
specifically focusing on the 
observed spectral 
index gradients, discrepancies in
Mach numbers 
between X-ray and radio measurements, 
and the role 
of pre-existing fossil plasma.

Two prominent scenarios -- 
direct acceleration of thermal
electrons in the ICM
 and re-acceleration
of fossil
electrons -- have been proposed to 
explain the 
origin of relativistic electrons in 
these shocks. 
In the acceleration framework,
shocks
directly 
accelerate ICM electrons to
relativistic energies 
from the thermal pool.
%To ensure consistency 
%between radio and X-ray observations, 
%the derived 
%Mach number from radio data 
%should closely match 
%that from X-ray measurements
%\citep[e.g.,][]%{2012A&ARv..20...54F,2012MNRAS.426...40B}.
However, the efficacy of direct electron 
acceleration in weak merger shocks 
is questioned due to the
limited ability
of weak shocks
to accelerates 
electrons to
relativistic energies. 
This challenge to the existence of
extended radio relics has been highlighted
in previous studies 
\citep[e.g.,][]{Kang_2012,2013MNRAS.435.1061P,2014IJMPD..2330007B,2016MNRAS.463.1534B,2016MNRAS.461.1302E}.
To address this discrepancy, 
an alternative re-acceleration
scenario has been proposed. 
In this scenario, 
low Mach number shocks re-energize a
pre-existing population of relativistic 
electrons rather than directly accelerating
thermal electrons.
These relic-associated electrons could
originate from nearby radio galaxies.
Thus, the Mach number measured from
radio could be different from that of 
measured using X-ray in re-acceleration scenario 
\citep{2014ApJ...785....1B,2015MNRAS.449.1486S}.

Work by \citet{2018MNRAS.478.2218H} 
showed spectral index maps of A1240
derived from observations spanning
the frequency range of 145 MHz to 3 GHz
(using LOFAR at 145 MHz, GMRT at 610 MHz, 
and JVLA at 2 -- 4 GHz).
Both radio relics exhibit spectral index 
($\alpha$) gradients that steepen toward the 
cluster center, reflecting electron cooling in the 
post-shock region of outward-traveling shock 
fronts. 
\citet{2018MNRAS.478.2218H} determined 
spectral indices of -0.94 $\pm$ 0.06 
and -0.97 $\pm$ 0.05 for the NW and SE shocks, 
respectively, corresponding to Mach numbers of
2.4 $\pm$ 0.1 and 2.3 $\pm$ 0.1. 
Remarkably, they identified two 
radio galaxies  with redshifts close to 
the 
cluster mean
redshift (redshifts $z$ =
0.193 and 0.192 -- each separated by less than 10 Mpc from A1240), close to the NE relic, which are potentially rich sources in 
supplying mildly relativistic electrons 
that contribute to synchrotron radio emission, and one 
at redshift $z$ = 0.152, which is close in projection to the SE relic but separated by more than 100 Mpc from A1240. 
%\textcolor{red}{
Peculiar velocities of approximately 500, 750, and 11,000 $\rm km~s^{-1}$ would explain the redshift differences, so we can rule out the radio source at 
$z$ = 0.152 as a cluster member.
Also, the velocity dispersion in a bound massive cluster stays within
a few thousand km/s.

The deep $\chandra$ 
observations allow us to measure 
Mach numbers more accurately from X-ray 
surface 
brightness profiles, providing a 
crucial comparison 
with radio-derived Mach numbers. 
We find significantly lower Mach numbers -- 
{ 1.41$^{+0.17}_{-0.19}$}
(for NW shock) 
and 1.49$^{+0.22}_{-0.24}$ (for SE shock) -- 
in contrast to the radio-derived values. 
{ Similar discrepancy between
X-ray and radio-derived Mach numbers 
have also been found in works by
\citet{2013ApJ...765...21S},
\citet{2015ApJ...812...49H},
\citet{2019ApJ...883..138R},
\citet{2021MNRAS.500..795D},
and \citet{2021MNRAS.506..396W}.}
%{ 
%\citet{2018MNRAS.478.2218H} showed
%that if the Mach numbers for both shocks
%in A1240 are $>$ 4.0, the required
%efficiency for accelerating 
%electrons is $<$ 1\%, suggesting
%the relativistic electrons observed 
%in the relics might be directly 
%accelerated from the thermal pool.
%Conversely, when the Mach number is 
%lower,
%specifically below 3, 
%it suggests a scenario of 
%re-acceleration of 
%pre-existing fossil electrons 
%from radio galaxies by the shock.}
%The lower
%Mach numbers of both shocks
%and the proximity of 
%radio galaxies to the relics 
%of A1240
%strongly support the 
%re-acceleration scenario over 
%direct acceleration 
%through DSA.
{
\citet{2021MNRAS.506..396W} showed 
that the 
radio observations tend to 
be sensitive to higher 
Mach numbers, whereas
X-ray observations yield an 
average Mach number for the relic 
as a whole. 
This distinction can potentially give
rise to variations in the Mach numbers 
deduced from the two observation methods
\citep{2022ApJ...927...80R}.
In addition,
a fraction of 
such disparities could be attributed to 
projection effects inherent 
in X-ray surface brightness analysis,
a factor that does not significantly 
influence radio observations. 
Abell~1240 is a double relic cluster, where
two relics are aligned on the sky plane.
This spatial arrangement 
implies that the merger plane
is very close to the plane
of the sky.

{
Here, we estimate the required
acceleration efficiency
($\eta$)
for both shocks, assuming the shocks produced the radio relics by accelerating electrons from the thermal pool. The required acceleration efficiency is defined as 
\begin{equation} \label{eq:effi}
    \eta = \frac{E_{\rm relic}}{\Delta F_{\rm KE}},
\end{equation}
where $E_{\rm relic}$ is the energy
flux of the accelerated relativistic 
electrons at the relic, defined
as, $E_{\rm relic}$ = $v_{\rm down}\ B^2/8\pi$, where $v_{\rm down}$ and 
$B$ are
the
the downstream velocity and magnetic 
field strength.
$\Delta F_{\rm KE}$ is the
available kinetic energy flux at the shock, 
given by
\begin{equation}
    \Delta F_{\rm KE} = 0.5 \rho_{\rm up} v^3_{\rm shock}(1 - \frac{1}{C^2}),
\end{equation}
where $\rho_{\rm up}$ and $v_{\rm shock}$ 
are the upstream
gas density and shock velocity, respectively, and  
$C$ is the density compression factor.
For the NW and SE shocks, we adopt 
pre-shock temperature of 2.3 keV
and an upstream electron density
of $\sim$ 10$^{-4}$ cm$^{-3}$, 
which
results to shock speeds of 
$v_{\rm shock}\ \sim$ 
910 km/s and 850 km/s.
Assuming equipartition of energy, the magnetic
field strength is $\sim$ 2.5~$\mu$G
at the both relic regions 
\citep{2009A&A...494..429B,2018MNRAS.478.2218H}. For the values presented above, Equation (\ref{eq:effi}) yields
a required acceleration efficiency of $\sim$ 1 for
both shocks. 
Achieving such high acceleration 
efficiency for such low Mach number
shocks is challenging for DSA-only 
acceleration
scenario. 

Our result strongly suggests 
that the NW and SE shocks in A1240 
did not primarily accelerate electrons 
from the thermal pool but rather 
re-accelerated pre-existing 
relativistic electrons.
A similar re-acceleration scenario has 
been proposed for A3411-3412, 
demonstrating the 
broad applicability of this mechanism 
\citep{2017NatAs...1E...5V}.
The remarkably high efficiency 
suggests 
that almost all of the kinetic
energy released during the shock 
goes into
the cosmic ray electrons and
the magnetic field. 
Consequently, there is little energy
left to increase the temperature
of thermal particles.
This means we do not anticipate a 
significant temperature 
rise across the shock front. 
%Additionally, this high efficiency 
%would impact the compression ratio
%within the shock, altering the
%connection between the density 
%and temperature jump.
However,
the substantial uncertainties 
in the
temperature measurements,
as seen in Figure 
\ref{fig:temp_profile},
leave plenty of latitude 
in this argument.
}

%\textcolor{red}{From Wittor et al. MNRAS 506, 396–414 (2021): ``(i) The high Mach number regions of the shock front are signifi- cantly more radio luminous. (ii) X-ray surface brightness analysis may suffer from projection effects, while radio measurements are basically unaffected by pro- jection. (iii) Weaker Mach numbers tend to be found in X-ray brighter regions.'' I think we should discuss that a bit.}

\section{Conclusion}{\label{sec:con}}
We have presented deep $\chandra$ observations
of the merging galaxy cluster A1240. 
Several previous observations in 
the radio band
have shown two extended radio relics in the
NW ($\sim$ 0.7 Mpc) 
and SE ($\sim$ 1.3 Mpc)
directions with $\sim$ 2 Mpc
separation between them, highlighting 
the energetic merging processes. 
With deep X-ray observations, we 
have measured the ICM properties across
both radio relics. 
Our measurements,
together with the 
radio observations,
provide a multi-wavelength
approach in understanding the complex
gas dynamics behind shock fronts.
We summarize our findings below-
\begin{itemize}
    \item Shock front edges can primarily be 
    located by using surface
    brightness profiles. We, therefore,
    have extracted
    high-quality surface brightness profiles 
    spanning both radio relics, as 
    depicted in Figure \ref{fig:image}.
    Both profiles are then fitted with 
    broken-powerlaw models to pinpoint 
    the surface brightness
    edge locations.
    We have found
    two distinct surface brightness edges
    located 
    at 1095 kpc in the 
    NW 
    direction and 1230 kpc 
    in the 
    SE direction from the X-ray center of
    A1240,
    notably coinciding with the locations of the 
    radio relics. 
    We have measured the 
    density jumps of $\rho_{\rm post}$/$\rho_{\rm pre}$ = 1.7 $\pm$ 0.3
    across the SE edge and 
    $\rho_{\rm post}$/$\rho_{\rm pre}$ = 1.6 $\pm$ 0.2
    across the NW edge.

    \item Shock fronts are characterized by
    sharp temperature jumps across the
    edge. To measure the temperatures across both surface brightness edges, 
    we extracted spectra from 
    the same
    sectors used for extracting
    surface brightness profiles in the NW and
    SE directions. Our ICM
    temperature 
    measurements showed 
    drops across
    both the NW ($\sim$ 1.6) and 
    SE ($\sim$ 1.7) edges
    ({ $\sim$ 2$\sigma$} level), 
    hinting the presence of
    the shock fronts associated 
    with the identified surface brightness edges. 
    Furthermore, a
    comparative 
    assessment was conducted by
    comparing 
    the temperature measurements 
    along the NW and SE directions 
    with those of
    E and W directions. 
    We found within 
    the central $\sim$ 700 kpc 
    the ICM temperatures in all directions 
    exhibited remarkable consistency, except for 
    the innermost bin in the E direction. 
    Conversely, the ICM temperatures in the
    regions just inside the SE and NW edges are
    found to be
     hotter at a $\sim$ 2$\sigma$ 
     level compared to 
    other two directions, further supporting
    the existence of shock-heated gas
    in the 
    NW and SE directions.

    \item To understand the electron-heating mode behind shocks, we compared the 
    Coulomb collisional model
    and instant equilibrium model predictions
    with the observed temperature profiles 
    across
    both shocks. 
    Figure \ref{fig:colli_vs_ins}
    illustrates the 
    comparison. 
    Though the temperatures
    of the post-shock gas for the NW and SE
    shocks favor the collisional model,
    we can not rule out either of the
    models, given the large measurement
    uncertainties in the gas 
    temperatures.

    \item Shock Mach numbers derived from X-ray
    observations in conjunction with the radio
    measurement provides a unique yardstick 
    in understanding long-debated electron
    acceleration mechanism by 
    the shocks.
    The $\chandra$ surface brightness
    profiles provide shock Mach numbers of
    $\cal{M}_{\rm SE}$ = 1.49$^{+0.22}_{-0.24}$ 
    and $\cal{M}_{\rm NW}$ = 1.41$^{+0.17}_{-0.19}$
    for the SE and NW directions, respectively,
    which are significantly
    different from 
    the Mach numbers measured
    from the profiles of
    the radio spectral index
    -- 
    $\cal{M}_{\rm SE}$ = 2.3 $\pm$ 0.1
    and 
    $\cal{M}_{\rm NW}$ =
    2.4 $\pm$ 0.1 (NW).
    {
    We estimated the required acceleration
    efficiency of $\sim$ 1 
    for both shocks. 
    Such high efficiency for such
    low Mach number shocks strongly
    indicates that the NW and SE shocks
    in A1240 re-accelerated pre-existing
    relativistic electrons rather
    directly accelerating them from 
    thermal pool.}

\end{itemize}

Our findings underscore the 
complexities of electron 
acceleration in weak merger shocks and 
emphasize the significance of 
re-acceleration scenarios.
Further investigations, combining 
multi-wavelength data and 
sophisticated modeling, are vital to 
unraveling the intricate mechanisms
governing particle acceleration in 
the dynamic environment of galaxy clusters.

\begin{acknowledgments}
We thank anonymous referee for providing
helpful comments and suggestions, which
improved the paper.
This work is based on observations
obtained with the $\chandra$ X-ray Observatory, a NASA mission. 
This paper employs a list of Chandra datasets, obtained by the Chandra X-ray Observatory, contained in~\dataset[DOI: 10.25574/cdc.190]{https://doi.org/10.25574/cdc.190}.
F.A.-S. acknowledges support from {\em Chandra} grant GO0-21119X. 
W.F. acknowledges support from
the
Smithsonian Institution, the 
Chandra High Resolution
Camera Project through 
NASA contract NAS8-03060,
and NASA Grants 80NSSC19K0116, 
GO1-22132X, and
GO9-20109X.
\end{acknowledgments}

\bibliography{sample631}{}
\bibliographystyle{aasjournal}

%%%%%%%%%%%%%%%%%%%%%%% individual images %%%%%%%%%%%%%%%%%%%%%%%%%%%%%%%%%%%%%%%%%%%

%%%%%%%%%%%%%%%%%%%%%%%%%%%%%%%%%%%%%%%%%%%%%%%%%%



%%%%%%%%%%%%%%%%%%%%%%%%%%%%%%%%%%%%%%%

%%%%%%%%%%%%%%%%%%%%%%%%%%%%%%%%%%%%%%%%%%%%%%%%%%

% Don't change these lines
\label{lastpage}

\end{document}